\documentclass{cernphprep}

\usepackage{graphicx}
\DeclareGraphicsExtensions{.eps,.pdf,.png,.gif,.jpg,.mht}
\usepackage{multirow}

\usepackage{amssymb}
\usepackage{verbatim}
\DeclareMathOperator{\nop}{\,}

\begin{document}
\renewcommand{\thefootnote}{\fnsymbol{\footnote}}
\begin{titlepage}
\PHnumber{2011--101}
\PHdate{ June 2011}
\EXPnumber{TOTEM 2011--01}
\EXPdate{4 July 2011}


\title{Proton-proton elastic scattering at the LHC energy of $\sqrt s =\,$ 7\,TeV}



\begin{Authlist}
The TOTEM Collaboration\\
G.~Antchev\footnote{INRNE-BAS, Institute for Nuclear Research and Nuclear Energy, Bulgarian Academy of Sciences, Sofia, Bulgaria}\addtocounter{footnote}{-1}, P.~Aspell$^{8}$, I.~Atanassov$^{8,}$\footnotemark, V.~Avati$^{8}$, J.~Baechler$^{8}$,
V.~Berardi$^{5b,5a}$, M.~Berretti$^{7b}$, M.~Bozzo$^{6b,6a}$,
E.~Br\"{u}cken$^{3a,3b}$, A.~Buzzo$^{6a}$, F.~Cafagna$^{5a}$, M.~Calicchio$^{5b,5a}$,
M.~G.~Catanesi$^{5a}$, C.~Covault$^{9}$, M.~Csan\'{a}d$^{4}$\footnote{Department of Atomic Physics, ELTE University, Hungary} , T.~Cs\"{o}rg\"{o}$^{4}$,
M.~Deile$^{8}$, E.~Dimovasili$^{8}$, M.~Doubek$^{1b}$,
K.~Eggert$^{9}$, V.Eremin\footnote{Ioffe Physical - Technical Institute of Russian Academy of Sciences}, F.~Ferro$^{6a}$, A. Fiergolski\footnote{Warsaw University of Technology, Poland}, F.~Garcia$^{3a}$, S.~Giani$^{8}$,
V.~Greco$^{7b,8}$, L.~Grzanka$^{8,}$\footnote{Institute of Nuclear Physics, Polish Academy of Science, Cracow, Poland}\addtocounter{footnote}{-2}, J.~Heino$^{3a}$, T.~Hilden$^{3a,3b}$,
M.~Janda$^{1b}$, J.~Ka\v{s}par$^{1a,8}$, J.~Kopal$^{1a,8}$, V.~Kundr\'{a}t$^{1a}$, K.~Kurvinen$^{3a}$,
S.~Lami$^{7a}$, G.~Latino$^{7b}$,
R.~Lauhakangas$^{3a}$, T. Leszko\footnotemark\addtocounter{footnote}{-3}
E.~Lippmaa$^{2}$,
M.~Lokaj\'{\i}\v{c}ek$^{1a}$, M.~Lo~Vetere$^{6b,6a}$, F.~Lucas~Rodr\'{\i}guez$^{8}$,
M.~Macr\'{\i}$^{6a}$, L.~Magaletti$^{5b,5a}$,
G.~Magazz\`{u}$^{7a}$, A.~Mercadante$^{5b,5a}$, M.~Meucci$^{7b}$,
S.~Minutoli$^{6a}$, F.~Nemes$^{4,}$\footnotemark\addtocounter{footnote}{3}
H.~Niewiadomski$^{8}$, E.~Noschis$^{8}$, T.~Novak$^{4,}$\footnote{KRF, Gy\"{o}ngy\"{o}s, Hungary},
E.~Oliveri$^{7b}$, F.~Oljemark$^{3a,3b}$, R.~Orava$^{3a,3b}$, M.~Oriunno$^{8}$\footnote{SLAC National Accelerator Laboratory, Stanford CA, USA},
K.~\"{O}sterberg$^{3a,3b}$, A.-L.~Perrot$^{8}$, P.~Palazzi$^{7b}$, E.~Pedreschi$^{7a}$,
J.~Pet\"{a}j\"{a}j\"{a}rvi$^{3a}$, J.~Proch\'{a}zka$^{1a}$, M.~Quinto$^{5a}$,
E.~Radermacher$^{8}$, E.~Radicioni$^{5a}$,
F.~Ravotti$^{8}$, E.~Robutti$^{6a}$,
L.~Ropelewski$^{8}$, G.~Ruggiero$^{8}$,
H.~Saarikko$^{3a,3b}$,  A.~Santroni$^{6b,6a}$,
A.~Scribano$^{7b}$, G.~Sette$^{6b,6a}$, W.~Snoeys$^{8}$, F.~Spinella$^{7a}$,
J.~Sziklai$^{4}$, C.~Taylor$^{9}$,
N.~Turini$^{7b}$, V.~Vacek$^{1b}$, M.~V\'{i}tek$^{1b}$, J.~Welti$^{3a,b}$, J.~Whitmore$^{10}$.\\
\vspace{0.5cm}
$^{1a}${Institute of Physics, Academy of Sciences of the Czech Republic, Praha, Czech Republic.}\\
$^{1b}${Czech Technical University, Praha, Czech Republic.}\\
$^{2}${National Institute of Chemical Physics and Biophysics NICPB, Tallinn, Estonia.}\\
$^{3a}${Helsinki Institute of Physics, Finland.}\\
$^{3b}${Department of Physics, University of Helsinki, Finland.}\\
$^{4}${MTA KFKI RMKI, Budapest, Hungary.}\\
$^{5a}${INFN Sezione di Bari, Italy.}\\
$^{5b}${Dipartimento Interateneo di Fisica di Bari, Italy.}\\
$^{6a}${Sezione INFN, Genova, Italy.}\\
$^{6b}${Universit\`{a} degli Studi di Genova, Italy.}\\
$^{7a}${INFN Sezione di Pisa, Italy.}\\
$^{7b}${Universit\`{a} degli Studi di Siena and Gruppo Collegato INFN di Siena, Italy.}\\
$^{8}${CERN, Geneva, Switzerland.}\\
$^{9}${Case Western Reserve University, Dept. of Physics, Cleveland, OH, USA.}\\
$^{10}${Penn State University, Dept. of Physics, University Park, PA, USA.}\\

\end{Authlist}
\vspace{2cm}
\begin{abstract}

Proton-proton elastic scattering has been measured by the TOTEM experiment at the CERN Large Hadron Collider at $\sqrt{s}$ = 7\,TeV in dedicated runs with the Roman Pot detectors placed as close as seven times the transverse beam size ($\sigma _{\rm beam}$) from the outgoing beams. After careful study of the accelerator optics and the detector alignment, $|t|$ , the square of four-momentum transferred in the elastic scattering process, has been determined with an uncertainty of $\delta t$ = 0.1\,\rm{GeV} $\sqrt{|t|}$.  In this letter, first results of the differential cross section are presented covering a $|t|$-range from 0.36 to 2.5\,GeV$^2$. The differential cross-section in the range 0.36 $< |t| <$ 0.47 GeV$^2$ is described by an exponential with a slope parameter $B = ( 23.6\pm 0.5_{\rm stat} \pm 0.4_{\rm syst}) \,\rm{GeV}^{-2}$, followed by a significant diffractive minimum at $|t| = (0.53 \pm 0.01_{\rm stat} \pm 0.01_{\rm syst})\,\rm{GeV}^2$. For $|t|$-values larger than $\sim 1.5\,\rm{GeV}^2$, the cross-section exhibits a power law behaviour with an exponent of -7.8~$\pm~0.3_{\rm stat} \pm 0.1_{\rm syst}$. When compared to predictions based on the different available models, the data show a strong discriminative power despite the small $t$-range covered.

\end{abstract}
PACS 13.60.Hb: Total and inclusive cross sections
\vspace{2cm}
\begin{center}
    {\em Accepted for publication in EPL}
\end{center}

\end{titlepage}








\section{Introduction}

Elastic $\rm {p p}$ and $\rm \bar{p} p$ scattering provides a sensitive probe for the structure of the proton, with a scale given by the impact parameter or, inversely,
by $t$ the four-momentum transfer squared. Increasing $|t|$ means looking deeper into the proton.
Depending on $t$, the amplitudes from different scattering processes contribute to the differential cross-section ${\rm d\sigma / d}t$: from Coulomb scattering at very small $|t|$ to non-perturbative nuclear scattering and -- after a particularly interesting transition range at intermediate $|t|$ -- to perturbative nuclear scattering at high $|t|$.
Early measurements~\cite{Bohm:1974tv,Kwak:1975yq} made almost 40 years ago at the Intersecting Storage Rings (ISR) at energies between 23\,GeV and 63\,GeV revealed a peculiar structure for ${\rm d\sigma / d}t$.
At low $|t| \sim (0.01$ -- $0.5)\,\rm GeV^{2}$ (already dominated by nuclear scattering), the differential $\rm {p p}$ cross-section showed an approximately exponential behaviour, ${\rm e}^{-B\,|t|}$, where $B$ is the slope parameter, followed by a diffractive minimum (the ``dip''), at $|t| \approx 1.4\,\rm GeV^{2}$, and subsequently a broad peak.
In $\rm \bar{p} p$ scattering~\cite{amos:1981dr,Ambrosio:1982zj,Breakstone:1984te,Amos:1985wx,Breakstone:1985pe} the dip is replaced by a shoulder at approximately the same $|t|$ position where the pp data have the minimum.
At larger $|t|$, ${\rm d\sigma / d}t$ for pp was found to decrease according to $|t|^{-n}$ with $n \approx 8$~\cite{Giacomelli:1979nu}.

Above the ISR energies, differential cross-section data have up to now only been available for $\rm \bar{p} p$ collisions from measurements at the S$\rm \bar{p} p$S collider~\cite{Arnison:1983mm,Bozzo:1985th,Bernard:1986ye}
and the Tevatron~\cite{Abe:1993xy,Amos:1992zn,Amos:1990fw,Brandt:2010zz} between $\sqrt{s} = 546\,\rm GeV$ and 1.96\,TeV, and showed the same structure as $\rm \bar{p} p$ data at the ISR.

The $\sqrt{s}$ dependence of the structure of ${\rm d\sigma / d}t$ exhibits two remarkable features: (1) the shrinkage of the elastic peak with increasing $\sqrt{s}$, manifest in an increase of the slope $B$~\cite{Amaldi:1978vc} and in the dip moving to lower $|t|$~\cite{Nagy:1978iw}, which
can also be interpreted
as an increase of the effective proton radius~\cite{Alberi:1981af}; and (2) the energy independence of the $|t|^{-8}$ power law at large $|t|$ as predicted by perturbative QCD~\cite{Donnachie:1979yu}.

The TOTEM experiment at CERN's Large Hadron Collider (LHC) is optimised for measuring elastic pp scattering over a $|t|$-interval ranging ultimately from
$10^{-3}$ to $10\,\rm GeV^{2}$ and thus offers an excellent opportunity to extend the earlier pp elastic scattering measurements to energies that are more than 100 times higher than the ISR energies where pp scattering was last studied.

TOTEM will also measure the total pp cross-section in dedicated special-optics runs, using the luminosity-independent method based on the Optical Theorem, and will study diffractive dissociation, including single, double and central diffractive topologies.

In this letter, we report the first measurement of elastic pp scattering in the $|t|$-range from 0.36 to 2.5\,GeV$^{2}$ using standard 2010 LHC beam optics with $\beta^{*} = 3.5\,$m. The extension to lower and larger $|t|$ will follow in later publications.

\section{The Roman Pot detectors}

To detect leading protons scattered at very small angles, silicon sensors are placed in movable beam-pipe insertions -- so-called ``Roman Pots" (RP) -- located symmetrically on either side of the LHC intersection point IP5 at distances of 215 -- 220\,m from the interaction point~\cite{Anelli:2008zza}. In order to maximize the experiment's acceptance for elastically scattered protons, the RP are able to approach the beam centre to a transverse distance as small as 1\,mm.

Each RP station is composed of two units separated by a distance of about 5\,m. A unit consists of 3 RPs, two approaching the outgoing beam vertically and one horizontally, allowing for a partial overlap between horizontal and vertical detectors.  The detectors in the horizontal pot complete the acceptance for diffractively scattered protons.

All RPs are rigidly fixed within the unit, together with a Beam Position Monitor (BPM).
One of the most delicate and difficult tasks of the experiment is to ensure the precision and the reproducibility of the alignment of all RP detector planes with respect to each other and to the position of the beam centre.

Each RP is equipped with a stack of 10 silicon strip detectors designed with the specific objective of reducing the insensitive area at the edge facing the beam to only a few tens of micrometers.
The 512 strips with 66\,$\mu$m pitch of each detector are oriented at an angle of $+ 45^\circ$ (five planes) and $- 45^\circ$ (five planes) with respect to the detector edge facing the beam.
A significant reduction of the uncorrelated background is achieved at the trigger level by implementing programmable coincidences requiring collinear hits in at least three of the five planes for each projection.
For this purpose radiation tolerant integrated circuits were mounted on the detector.
The 10 detector planes within a stack were aligned and mounted with a precision of 20\,$\mu $m.
Then the three RPs and the BPM were surveyed. The movements of the RPs via step motors (5\,$\mu $m step) are independently verified with displacement inductive sensors (LVDT) with 10\,$\mu $m precision.
During the measurement the detectors in the horizontal RPs overlap with the ones in the vertical RPs, enabling a precise (10\,$\mu $m) relative alignment of all three RPs in a unit by correlating their positions via common particle tracks.
A dedicated beam fill is used to align all the RPs symmetrically with respect to the beam centre by moving them against the sharp beam edge cut by the beam collimators until a spike of beam losses was recorded downstream of the RPs. The precision of this procedure is determined by the movement step size and amounts to about 50\,$\mu$m for the alignment in 2010.

\begin{figure}[ht]
\begin{center}
\includegraphics[width=0.45\textwidth]{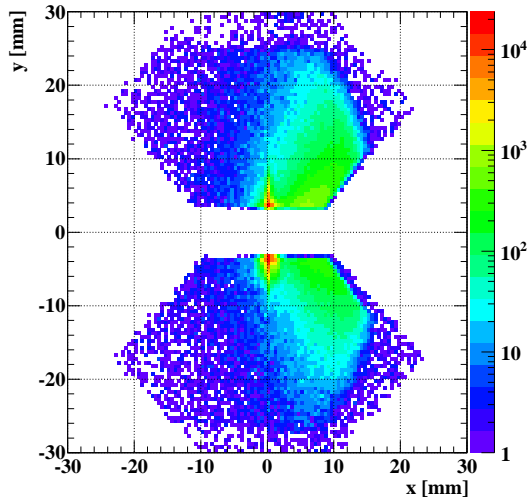}
\end{center}
\caption{The intercepts of all selected reconstructed tracks (see text) in a
scoring plane transverse to the beam at 220\,m.} 
\label{fig_reco-tracks}
\end{figure}

In a station, the duplication of the RP units with a long lever arm between the near and the far units has several important advantages.
First, the local track angles in the x- and y- planes perpendicular to the beam are reconstructed with a precision of 5 to 10\,$\mu$rad, helping to identify the background. These angles are related via the beam optics to the scattering angle of the proton at the vertex. For the standard beam optics, this relation offers the only way to measure the horizontal scattering angle ($\Theta^* _x$) with good precision.
Second, the proton trigger selection by track angle uses both units independently, resulting in a high trigger efficiency of $(99\pm1 )\%$.

\section{Data selection and analysis}

The data presented here were taken with the standard LHC 2010 optics ($\beta ^*$  = 3.5 m) during a TOTEM dedicated run with four proton bunches of nominal population ($7 \times 10^{10}$ p/bunch) per beam with a total integrated luminosity of 6.1\,nb$^{-1}$.  This low-luminosity configuration allowed the detectors to approach the beams to a distance as small as 7 times the transverse beam size $\sigma_{\rm{beam}}$.

A reconstructed track in both projections in the near and in the far vertical RP unit is required on each side of the IP.
The two diagonals {\em top left of IP -- bottom right of IP} and {\em bottom left of IP -- top right of IP}, tagging possible elastic candidates, are used as almost independent experiments with slightly different optics corrections, yet constrained by the alignement of the RPs.


\begin{figure}[!h]
\begin{center}
\includegraphics[width=0.45\textwidth]{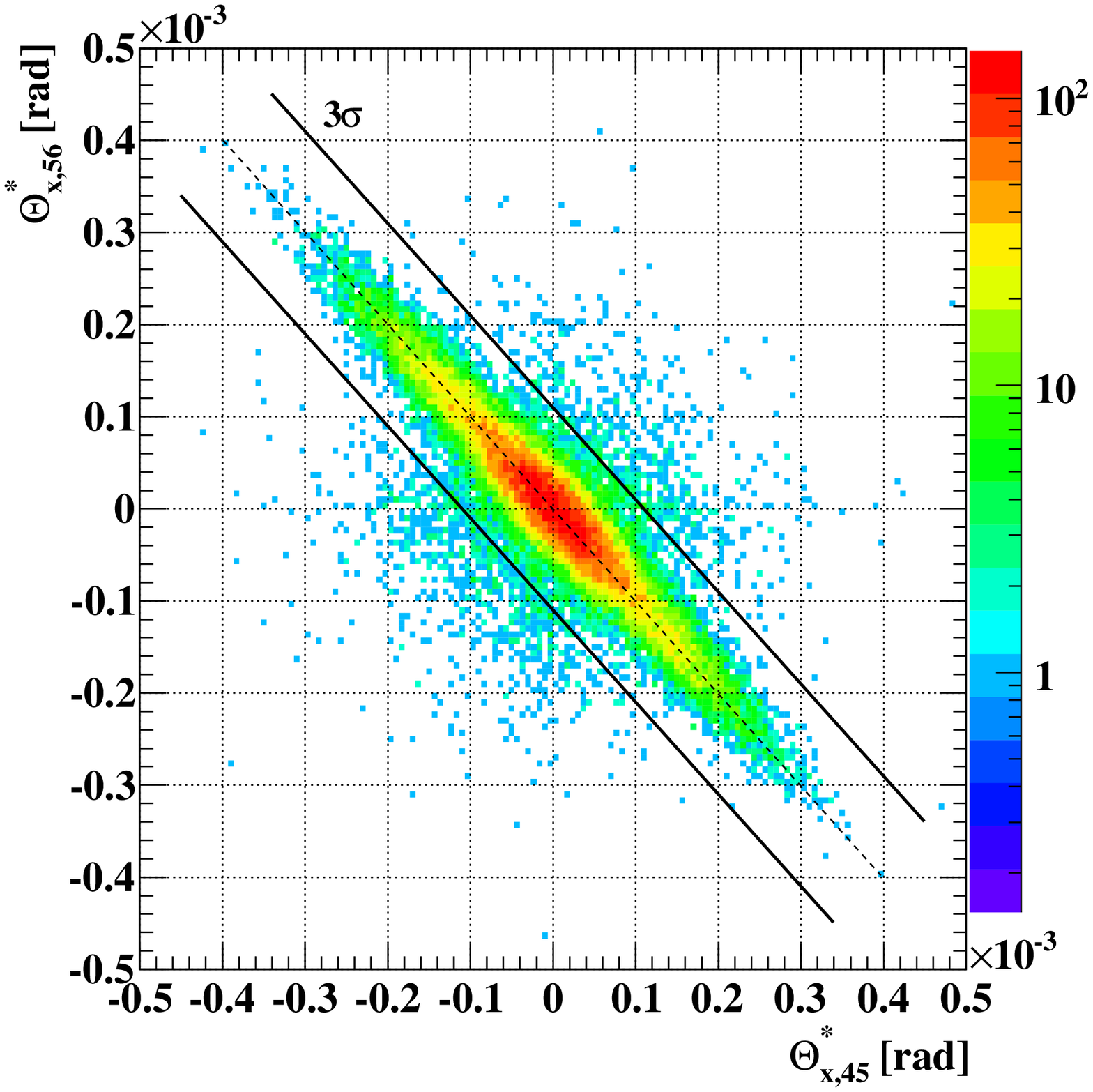}
\hfill
\includegraphics[width=0.45\textwidth]{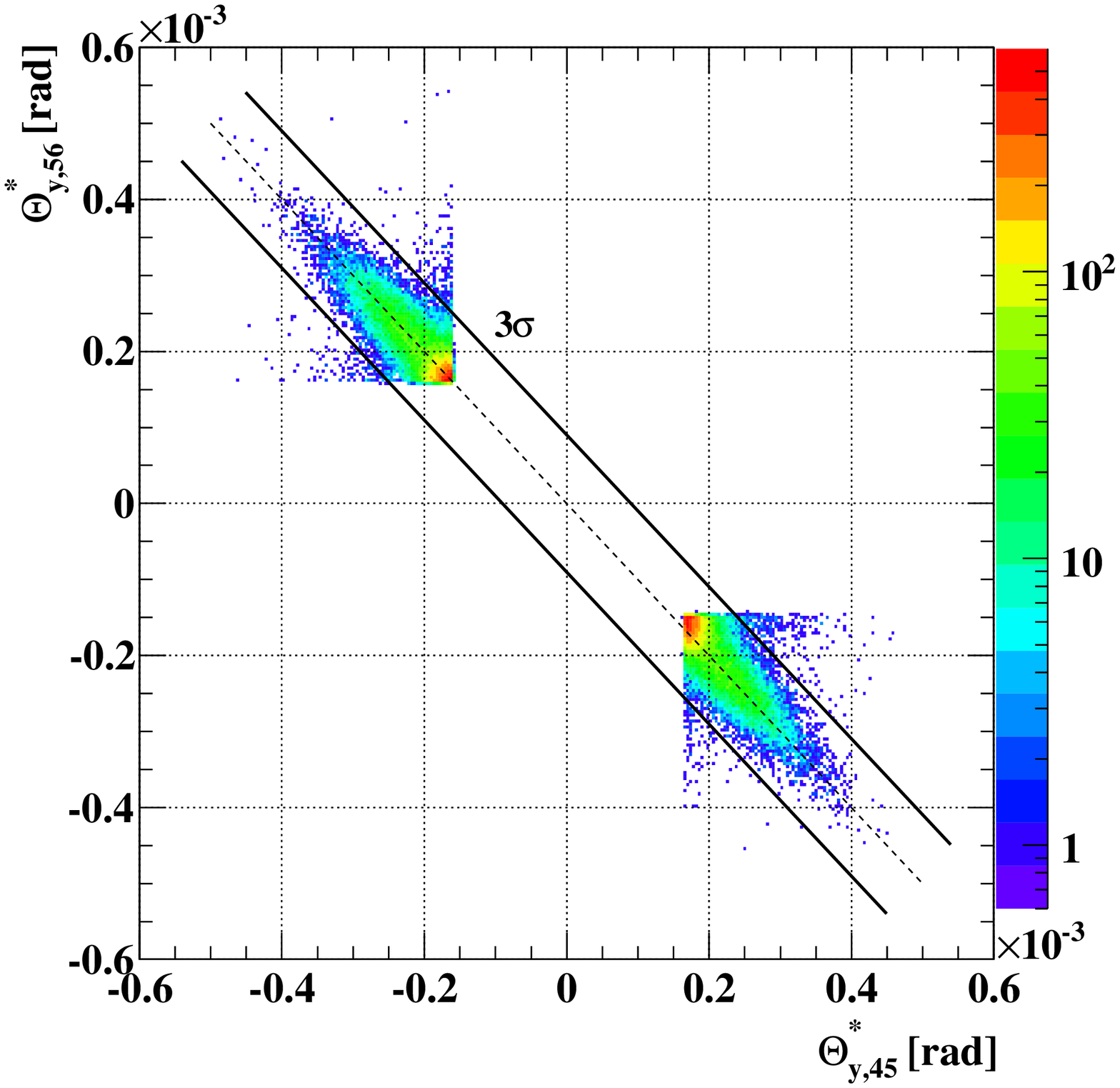}
\end{center}
\caption{The correlation between the reconstructed proton scattering angles $\Theta^* _y$ (top plot) and $\Theta^* _x$ (bottom plot) on both sides of the IP (``45" = left of IP5, ``56" = right of IP5).  The observed spread is in agreement with the beam divergence.} 
\label{fig_angle-correl}
\end{figure}

\begin{table}[!htb]
\begin{center}
\caption{Event sample reduction by the analysis cuts.}
\begin{tabular}{|l c|}\hline
 Total triggers& 5.28 M \\
\hline
Reconstructed tracks \& elastic topology & 293 k \\
\hline
Low $|\xi |$ selection & 70.2 k \\
\hline
Collinearity cuts & 66.0 k \\
\hline
\end{tabular}
\label{tab_dataflow}
\end{center}
\end{table}
The intercepts of the selected tracks in a scoring plane at 220\,m transverse to the beam direction are shown in Fig.~\ref{fig_reco-tracks}. In first approximation prior to refined optical corrections, the displacement along the y-axis is proportional to the vertical scattering angle, whereas for the present standard LHC optics the horizontal scattering angle does not lead to a sizeable horizontal displacement. However, protons with a momentum loss $\xi  = \Delta p / p$ are shifted in the positive x direction by the amount $x =\xi D$ ($D$ is the dispersion).
\begin{table*}[!hbt]
\begin{center}
\caption{Analysis Corrections and Systematics {\scriptsize($\delta$ denotes an uncertainty; the optical function $L$ relates the displacement to the scattering angle: $x,y = L_{x,y}\;\Theta^*_{x,y}$; $\Delta s \; (\approx 5\,{\rm m})$ is the distance between the two units in one RP station).}} 
\resizebox{17.5cm}{!}{
\renewcommand{\arraystretch}{1.5}
\begin{tabular}{|l|c|c|c|c|}
\hline


Correction & Effect on & Functional form & Total values or integral & Details\\ \hline

Recorded & \multirow{3}{*}{d$\sigma$/d$t$} &const($t$) & Efficiency-corrected int. Luminosity & Int. Luminosity $(6.1 \pm 0.2) \rm{nb}^{-1} $  \\
\multirow{2}{*}{Luminosity} &&\multirow{2}{*}{mult. factor}&\multirow{2}{*}{$(6.03 \pm 0.36)\, \rm{nb}^{-1} $ }& Trigger eff. $ (99 \pm 1)\,\% $\\
&& && DAQ eff. $ (99 \pm 1)\, \%$ \\ \hline

\multirow{2}{*}{Inefficiency} & \multirow{2}{*}{d$\sigma$/d$t$} & Ineff. = const($t$) & \multirow{2}{*}{Tot. ineff. = $(30 \pm 10) \% $ }& Detector $1 \% $ \\
&& mult. corr. factor = (1 + ineff.) && Event reconstruction $(29 \pm 10 )\%$ \\ \hline

\multirow{4}{*}{Acceptance} & \multirow{4}{*}{d$\sigma$/d$t$} & \multirow{4}{*}{$ \begin{array} {c} \text{Hyperbola function:} \\ f_A\approx 1.3 + \frac {0.3}{(|t|-0.3)}  \\ \text{ mult. corr. factor} \end{array} $ } & \multirow{4}{*}{$f_A= \left \{ \begin{array} {c} 4.96 ~ \pm 0.05 \vert_{|t|=0.4 \,\rm{GeV}^2} \\  2.92 ~ \pm 0.03 \vert_{|t|=0.5\, \rm{GeV}^2} \\1.55 ~ \pm 0.02 \vert_{|t|=1.5 \,\rm{GeV}^2} \end{array} \right. $ } & \multirow{4}{*}{$ \begin{array} {r|r}  y :~2.2 \vert_{|t|=0.36 \, \rm{GeV}^2} & \phi:~4.5 \vert_{|t|=0.36 \,\rm{GeV}^2} \\ 1.5 \vert_{|t|=0.4 \,\rm{GeV}^2}\; & 3.5 \vert_{|t|=0.4 \,\rm{GeV}^2} \;\\ 1.1 \vert_{|t|=0.5 \,\rm{GeV}^2} \;& 2.6 \vert_{|t|=0.5 \,\rm{GeV}^2} \;\\ 1.0 \vert_{|t|=1.5 \,\rm{GeV}^2} \;& 1.5\vert_{|t|=1.5 \,\rm{GeV}^2}\; \end{array}$ } \\
&&&& \\
&&&& \\
&&&& \\ \hline

\multirow{3}{*}{Background} & \multirow{3}{*}{d$\sigma$/d$t$} & Parameterisation & \multirow{3}{*}{$\frac{\int {\rm{bkg.}}\; {\rm d}t}{{\rm total}}=(8 \pm 1) \%$ }& \multirow{3}{*}{$ \frac {\rm{bkg.}} {{\rm total}} = \left \{ \begin{array} {c}(11 \pm 2) \% \vert_{|t|=0.4 \rm{GeV}^2} \\ (19 \pm 3) \% \vert_{|t|=0.5 \rm{GeV}^2} \\ (0.8 \pm 0.3) \% \vert_{|t|=1.5 \rm{GeV}^2} \end{array} \right . $ } \\
&& $  \rm{bkg.} = 1.16 \, {\rm e} ^{-6.0|t|}$ &&  \\
&& mult. corr. factor $ = ( 1 -\frac{\rm bkg.}{\rm total} ) $ & & \\ \hline

& \multirow{4}{*}{$t \rightarrow {\rm d}\sigma$/d$t$} & \multirow{2}{*}{$f_u(\Theta^*)=\frac {\rm{unsmeared}}{\rm{measured}} $} & \multirow{4}{*}{$f_u= \left \{ \begin{array} {c}
{0.55}\, {\scriptstyle\sideset {}{_{- 0.09}^{+0.02}}\nop } \vert_{|t|= 0.36 \rm{GeV}^2, \Theta =170 \mu \rm{rad}}\\
{0.51}\, {\scriptstyle\sideset {}{_{- 0.10}^{+0.02}}\nop } \vert_{|t|= 0.4 \rm{GeV}^2, \Theta =181 \mu \rm{rad}}\\
{0.54}\, {\scriptstyle\sideset {}{_{- 0.15}^{+0.04}}\nop } \vert_{|t| =0.5 \rm{GeV}^2, \Theta =202 \mu \rm{rad}}\\
{0.91}\, {\scriptstyle\sideset {}{_{- 0.13}^{+0.10}}\nop } \vert_{|t| =1.50 \rm{GeV}^2, \Theta =350 \mu \rm{rad}} \end{array} \right .  $ } & \multirow{2}{*}{Dominant contribution}\\
Resolution && & & \\
unfolding & & \multirow{2}{*}{ mult. corr. factor } & & ${ \delta \Theta ^* =\frac {\rm{Beam ~divergence}}{\sqrt{2}} } = {\scriptstyle{12-13~ \mu \rm{rad}}}$ \\
&&& & \\ \hline

\multirow{3}{*}{Alignment} & \multirow{3}{*}{$t$} & \multirow{3}{*}{$\delta t_x=2p/({ \Delta s \;{\rm d} L_x / {\rm d} s })\sqrt{|t_x|}\, \delta x $ }&  $\delta t/t = 0.6 \% \vert_{|t|=0.4 \rm{GeV}^2}$ & Track based alignment for 2\\
& & & & mechanically constrained diagonals:\\
& &$\delta t_y=2p/L_y \sqrt{|t_y|}\, \delta y $ & $\delta t/t = 0.3 \% \vert_{|t|=1.5 \rm{GeV}^2}$ & $ \delta x <  10 \mu {\rm m};\;\; \delta y = 10 \mu {\rm m} $ \\ \hline

\multirow{4}{*}{Optics} & \multirow{4}{*}{ $t$ }& $t_x = f(k,\psi,p) ; \, \, t_y = f(k,\psi,p)$ & $\frac {\delta({\rm{d}} L_x / {\rm{d}}s)}{{\rm{d}} L_x / {\rm{d}}s} = 1 \% $ & ${{\delta k} \over k} = 0.1 \% $ \\
& & $k$: magnet strength & $\frac{\delta L_y}{ L_y} = 1.5 \% $ & $ {\delta \psi}  = 1 \rm{mrad}$ \\
& & $\psi$: magnet rotation & ${{\delta t}\over {t}} = 2 \%$ & $ {{\delta p} \over {p}} = 10^{-3}$ \\
& & $p$: LHC beam momentum & &  \\ \hline

\end{tabular}
}
\label{tab_resol-corrections}
\end{center}
\end{table*}
Elastically scattered protons are therefore recorded close to $x = 0$ while diffractive protons have positive x values due to $D$.
An accumulation of the elastically scattered protons at the detector edge and close to $x = 0$ is clearly visible in the raw distribution of Fig.~\ref{fig_reco-tracks}.
Hence a low $\xi$ requirement ($|x| < 0.4\, {\rm mm}$) is the first criterion for selecting elastic candidate events.

\begin{table*}[!tb]
\begin{center}

\caption{Statistical errors and systematic uncertainties on $t$ and ${{\rm{d} \sigma / \rm{d}t}}$. } 
\renewcommand{\arraystretch}{1.8}
\begin{tabular}{c|c|c|c|}
\cline{2-4}
& ${\delta t \over t }$ on single $t$ meas. & $\delta t = \delta\sideset{}{_{t}^{\rm stat}}\nop (t)\oplus {\delta}_{t}^{\rm syst} (t)$ & $ \delta( {\rm{d} \sigma / \rm{d}t}) = \delta\sideset{}{_{{\rm{d} \delta / \rm{d}t}}^{\rm stat}} \nop(t) \oplus {\delta} _{\rm{d} \sigma / \rm{d}t}^{\rm syst} (t)$ \\
\cline{1-4}
\multicolumn{1}{|c|}{$|t| = 0.4 \, \rm{GeV}^2 $} & 13 \% {\scriptsize(from beam div.)} & ${\delta t \over t }= \pm 0.5 \%^{\rm stat} \pm 2.6 \%^{\rm syst}$ & ${\delta({\rm{d} \sigma / \rm{d}t})\over{{\rm{d} \sigma / \rm{d}t}}}= \pm 2.6 \%^{\rm stat} \sideset {}{_{- 37}^{+25}}\nop \%^{\rm syst}$\\ \cline{1-4}
\multicolumn{1}{|c|} {$|t| = 0.5 \, \rm{GeV}^2 $ }& 12 \%  {\scriptsize(from beam div.)} & ${\delta t \over t} = \pm 0.7 \%^{\rm stat} \pm 2.5 \%^{\rm syst}$ & ${\delta({\rm{d} \sigma / \rm{d}t})\over{{\rm{d} \sigma / \rm{d}t}}}= \pm 4.4 \%^{\rm stat} \sideset {}{_{-39}^{+28}\nop \%^{\rm syst}} \nop  $\\ \cline{1-4}
\multicolumn{1}{|c|} {$|t| = 1.5 \, \rm{GeV}^2 $} & 7 \%  {\scriptsize(from beam div.)}  & ${\delta t \over t} = \pm 0.8 \%^{\rm stat} \pm 2.3 \%^{\rm syst}$ & ${\delta({\rm{d} \sigma / \rm{d}t})\over{{\rm{d} \sigma / \rm{d}t}}}= \pm 8.2 \%^{\rm stat} \sideset {}{_{-30}^{+27}}\nop \%^{\rm syst}$\\ \cline{1-4}
\end{tabular}
\label{tab_errors}
\end{center}
\end{table*}

Using the optical functions, the vertical  ($\Theta^* _y$) and horizontal  ($\Theta^* _x$) scattering angles are deduced from the measurements at the RP stations, $\Theta^* _y$ from the track displacement in y, with the minimum angle determined by the closest detector approach to the beam, and $\Theta^* _x$ from the track angle at the RP stations.
As a consequence of the collinearity of the elastically scattered protons, the angles  $\Theta^* _y$ and  $\Theta^* _x$ should be the same on both sides of IP5 except for small fluctuations arising from the beam divergence. Fig.~\ref{fig_angle-correl} demonstrates the impressive correlation between the scattering angles on both sides with a spread in agreement with the beam divergence, and from which a t-resolution of $\delta t$ = 0.1\,\rm{GeV} $\sqrt{|t|}$ has been deduced using the relation $t = - p^2 \Theta^{*\,2}$.


The collinearity of the elastically scattered protons and the interplay of the two diagonals was also used to fine-tune the LHC optics within the allowed limits of the magnet rotations and strengths. The fact that one unique set of magnet tunes leads to an agreement between data from both diagonals, gives confidence in the method.

Collinearity cuts at 3 $\sigma$ (drawn in the scattering angle correlation plots in Fig.~\ref{fig_angle-correl}) were applied to reduce the final background.
Table~\ref{tab_dataflow} summarizes the flow of the data reduction to the final sample of elastic candidates for the sequence of cuts applied.
Great care was taken to understand the factor of about 20 between the triggered and the analysed events, by studying the topological patterns and by visual event scans.
Table~\ref{tab_resol-corrections} gives a quantitative overview of the analysis corrections and associated systematics. Details of the analysis methods will be extensively addressed in a forthcoming TOTEM paper.


\begin{figure*}[!hb]
\begin{center}
\includegraphics[width=0.90\textwidth]{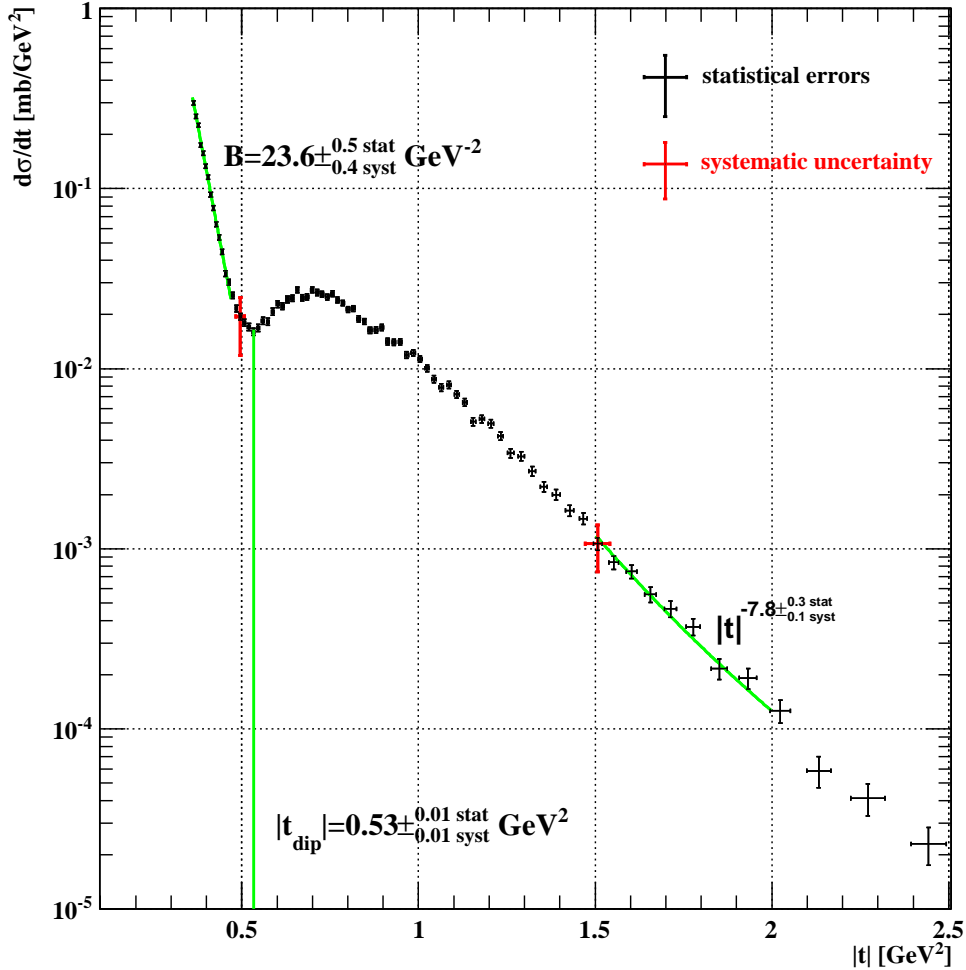}
\end{center}
\caption{The measured differential cross-section $\rm{d}\sigma/\rm{d}t$. The superimposed fits and their parameter values are discussed in the text.} 
\label{fig_t-distribution}
\end{figure*}

\begin{table*}[!ht]
\begin{center}
\caption{The values of the elastic slope parameter $B({|t| = 0.4 ~{\rm GeV}^2})$, the $|t|$-position of the diffractive minimum, $|t_{\rm dip}|$, the exponent of the power law behaviour at large $|t|$, and the diffraction cross-section at $|t|=0.7{\rm GeV}^{2}$, as extracted from the prediction of several models and compared to the measured values of the same quantity.} 
\renewcommand{\arraystretch}{1.4}  
\scalebox{0.85}{
\begin{tabular}{|c|c|c|c|c|}\hline 
\multirow{2}{*} {Models' prediction} & $ B\scriptstyle{(|t| = 0.4 ~{\rm GeV}^2)}$ & $|t_{\rm dip}| $ & $n$ in $ |t|^{-n}\scriptstyle{ (1.5 - 2\,\rm{GeV}^2)}$ & {${\rm d} \sigma/ {\rm d} t \scriptstyle{(|t| = 0.7 ~{\rm GeV}^2)} $ } \\
 &$\scriptstyle{\rm{[GeV}^{-2}{\rm ]}}$ & $\scriptstyle{[{\rm GeV}^2 {\rm ]}} $ & &$\scriptstyle {\rm{[mb/GeV}^{2}{\rm ]}} $\\ \hline
M.M. Block et al.~\cite{Block:2011uy}       & 24.4 & 0.48 & 10.4 & $9.1 \,\cdot 10^{-2}$ \\
C. Bourrely et al.~\cite{Bourrely:2002wr}   & 21.7 & 0.54 & 8.4 & $4.8 \,\cdot 10^{-2} $\\
M.M. Islam et al.~\cite{Islam:2009zz}       & 19.9 & 0.65 & 5.0 & $8.2 \,\cdot 10^{-3} $ \\
L.L. Jenkovszky et al.~\cite{Jenkovszky:2011hu} & 20.1 & 0.72 & 4.2 & $6.1 \,\cdot 10^{-3} $ \\
V.A. Petrov et al.~\cite{Petrov:2002nt}     & 22.7 & 0.52 & 7.0 & $4.6 \,\cdot 10^{-2} $ \\ \hline
This measurement                    & $23.6 \;{\pm0.5^{\rm stat}}\;{ \pm0.4^{\rm syst}}$ & $0.53 \;{\pm0.01^{\rm stat}} \;{\pm0.01^{\rm syst}} $ & $7.8\; {\pm0.3^{\rm stat}}\; { \pm0.1^{\rm syst}} $ & $ 2.7 \, \cdot 10^{-2}\; {+3.7\%^{\rm stat}}\; ^{ +26\%^{\rm syst}}_{-21\%^{\rm syst}}$  \\ \hline
\end{tabular}
}
\label{tab_model}
\end{center}
\end{table*}

The time dependent instantaneous luminosity was taken from the CMS measurement~\cite{int:lumi1,int:lumi2}.
Its determination is based on a van der Meer scan whose uncertainty was $4\,\%$ for the data presented in this paper.
The recorded luminosity is derived by integrating the luminosity, the trigger efficiency and the DAQ efficiency over all the different runs taken.


The inefficiency of the silicon detectors is very small. However, due to the inability to reconstruct multiple tracks, protons cannot be reconstructed if they produced showers.
An average inefficiency of 3 -- 7 \% per pot and tracks induced correlations lead to an event reconstruction inefficiency of $(29 \pm 10)\%$ (determined from the data).
The pile-up of minimum-bias onto elastic events is $ < 0.5 \% $ .

The total acceptance has been computed as a function of the vertical direction $y$ and the azimuth $\phi$.
The correction factors are large at $t$ values close to the acceptance limits at the detector edges.
The beam divergence causes large acceptance losses in $y$ near the detector edges, corresponding to $t$ values below $0.5 \, {\rm GeV}^2$.
The straight detector edges cause a strongly $t$-dependent acceptance in $\phi$.
To avoid large acceptance corrections and thus significant systematical uncertainties on ${{\rm{d} \sigma / \rm{d}t}}$ the present measurements have been limited to $0.36 \, <|t| < 2.5 \,\rm{GeV}^2$.
A $t$-dependent background of $(8\pm 1 )\%$ in the elastic candidates sample inside the selection cuts was evaluated from the data studying events lying outside the cuts.

The resolution effects and bin migration due to the beam divergence have been unfolded analytically and with a Monte Carlo method, obtaining consistent results.
These results have been confirmed by comparing the unfolded distribution with an uncorrected distribution sample not suffering from resolution effects (i.e. selected within the tight limits of $\pm 0.5 \sigma$ of the beam divergence).

The alignment of the RPs has been optimized by reconstructing parallel tracks going through the overlap between vertical and horizontal RPs, both in the ``near" and ``far" units.
The horizontal alignment with respect to the beam centre has been achieved with the help of low $\xi$ proton tracks, the vertical by matching the $\Theta^* _y$ distributions of elastic protons detected by the two diagonals. The final uncertainty is less than $10 \mu {\rm m}$.

The relevant optics parameters, such as magnet strengths, magnet rotations, and the LHC absolute momentum scale have been obtained by refitting the data within their nominal uncertainties, obtaining reasonable pull distributions.


Table~\ref{tab_errors} contains the  propagated contributions to $t$ and to d$\sigma$/d$t$ from all statistical and systematic uncertainties for different values of $t$.
The statistical error in $t$ is given by the beam divergence whereas the statistical error in d$\sigma$/d$t$ is given by the number of events.
The systematic uncertainty in $t$ is dominated by optics and alignment.
The systematic uncertainties in d$\sigma$/d$t$ are dominated by the uncertainty on the efficiency correction (t-independent) and on the resolution unfolding, which depends on the $t$ measurement errors and hence on the uncertainty on the beam divergence.
Both systematic uncertainties are correlated in $t$, therefore they mainly represent a global shift of the absolute scale of the d$\sigma$/d$t$ distribution.


\begin{figure*}[!htb]
\begin{center}
\includegraphics[width=.80\textwidth]{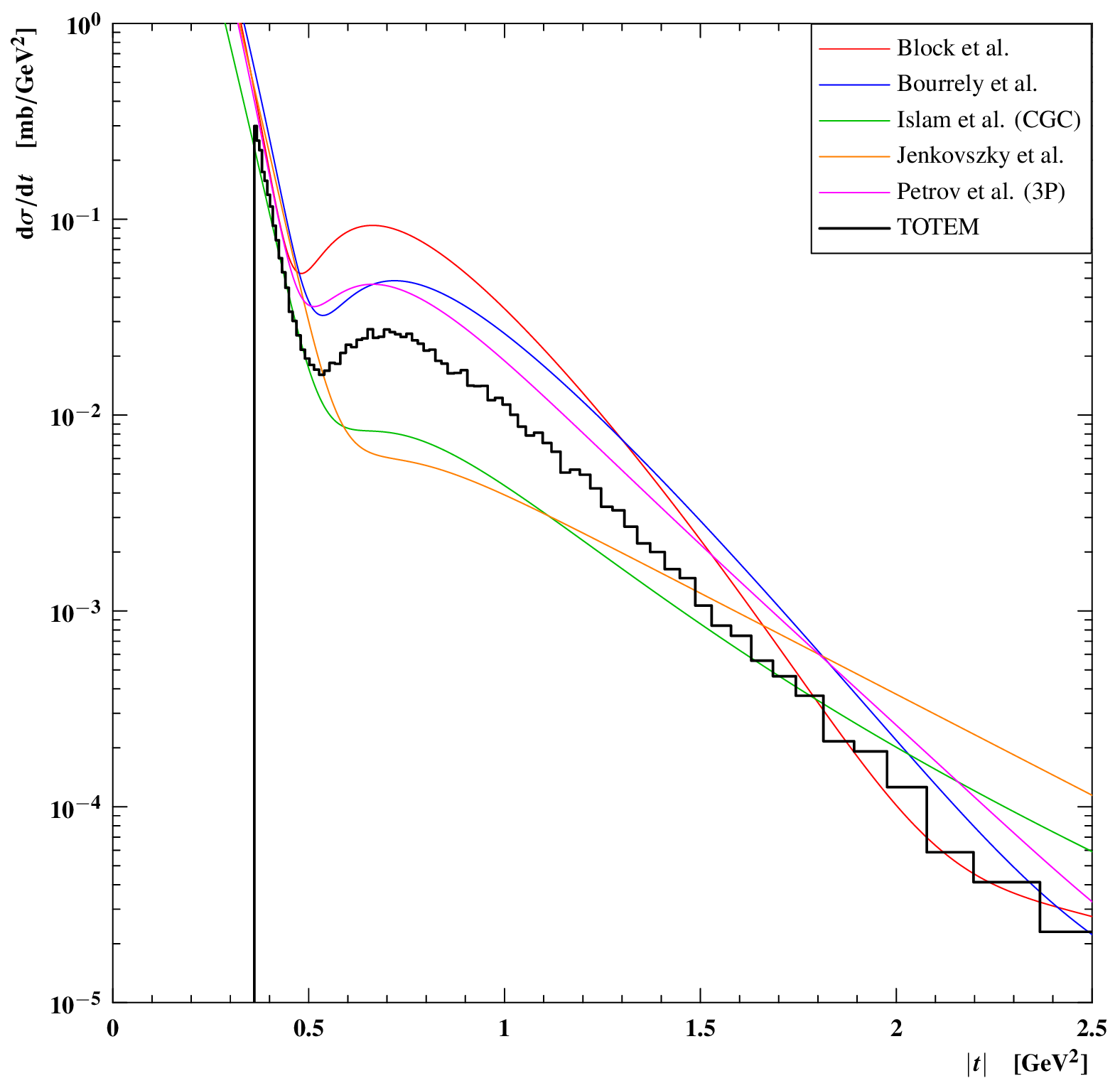}
\end{center}

\caption{The measured ${\rm d} \sigma/ {\rm d} t$ compared to the predictions of several models (see Table~\ref{tab_model}).} 
\label{fig_model}
\end{figure*}

\section{The differential cross-section}

After unfolding and inclusion of all systematic uncertainties (Tables~\ref{tab_resol-corrections} and~\ref{tab_errors}), the final differential cross-section $\rm{d}\sigma/\rm{d}t$ for elastic pp scattering is given in Fig.~\ref{fig_t-distribution} covering a $|t|$ range from 0.36 to 2.5 \,GeV$^2$. At $|t|$ values below 0.47\,GeV$^2$, the data can be described by an exponential function proportional to ${\rm e}^{-B|t|} $  with slope $B=(23.6 \pm 0.5_{\rm stat} \pm 0.4_{\rm syst})\,{\rm GeV}^{-2}$.
$B$ is expected to change at smaller $|t|$ values and differs considerably between the models considered.
The approximately exponential behaviour is followed by a diffractive minimum at $|t| = (0.53  \pm 0.01_{\rm stat} \pm 0.01_{\rm syst})\,{\rm GeV}^2$. This pronounced dip, observed in pp but not in $ {\bar {\rm p}} p$ scattering, moves to smaller $|t|$ values with increasing collision energy. This trend already observed at the ISR is now confirmed at $\sqrt{\rm s}$ = 7\,TeV.
Above the dip structure the differential cross-section becomes flatter and can be described with a power law $|t|^{-n}$ with an exponent $n= 7.8 \pm 0.3_{\rm stat} \pm 0.1_{\rm syst}$ for $|t|$-values between 1.5\,GeV$^2$ and 2.0\,GeV$^2$.

\section{Model comparison}

In Fig.~\ref{fig_model} the measured differential cross-section $\rm{d}\sigma/\rm{d}t$ is compared to the predictions from several models~\cite{Block:2011uy, Bourrely:2002wr,Islam:2009zz, Jenkovszky:2011hu, Petrov:2002nt} at $\sqrt{s}$~=~7\,TeV~\cite{Kaspar:2011zz}.
The extracted slope parameter $B(|t| = 0.4\,{\rm GeV}^2)$, the $|t|$-position of the diffractive minimum, $|t_{\rm dip}|$, the exponent $n$ at large $t$ and the differential cross-section at $|t|=0.7{\rm GeV}^{2}$ are listed in Table~\ref{tab_model} for a quantitative comparison.



Two models~\cite{Bourrely:2002wr, Petrov:2002nt} are consistent with the data for the slope parameter $B$ at $|t|=$ 0.4\,{\rm GeV}$^2$, the dip position, $|t_{\rm dip}|$, and the exponent $n$ at large $|t|$, but they both disagree with the cross-section in the measured range. The other three models~\cite{Islam:2009zz, Block:2011uy, Jenkovszky:2011hu} are less consistent with the data presented here.

\section{Outlook}

For further understanding of pp elastic scattering the $|t|$-range has to be considerably extended. The development of the approximately exponential behaviour at low $|t|$ is fundamental for the extrapolation to the Optical Point at $t$ = 0 and hence for the measurement of the elastic scattering and the total cross-section. In the large $|t|$ region, where a better understanding of the interplay between non-perturbative and short-distance perturbative-QCD dynamics is required, the cross-sections of the different models vary over orders of magnitude.

During 2010, TOTEM has already accumulated statistics ( 5.8\,pb$^{-1}$) in high-luminosity runs to extend the $|t|$ range to $\sim$4 -- 5\,GeV$^2$. The large $\beta ^*$~=~90\,m optics was developed and successfully tested in spring 2011 and first special fills are expected soon to reach $|t|$-values for elastic pp scattering around $10^{-2}$\,GeV$^2$. Total cross-section measurements are then expected together with extensive studies of diffractive phenomena.

\begin{center}
$\ast\ast\ast$
\end{center}

We thank our support staff and the members of various CERN groups (M.~Battistin, C.~Bault, S.~Blanchard, F.~Cossey Puget, E.~David, A.~de Oliveira, M.~Dupont, M.~Dutour, B.~Farnham, P.~Guglielmini, A.~Guipet, B.~Henrist, C.~Joram, L.-J.~Kottelat, C.~Lasseur, D.~Macina, I.~McGill, J.~No\"{e}l, A.~Numminen, X.~Pons, S.~Rangod, S.~Ravat, A.~ Thys, E.~Tsesmelis, M.~Van~Stenis) for contributing to the construction, installation and commissioning of the detector systems.

We are grateful to:

- the PH-ESE, TE-MPE groups (in particular R.~De~Oliveira) for the production of electronics components;

- V.~Boccone, S.~Cerchi, R.~Cereseto, S.~Cuneo, M.~Marchelli, A.~Morelli, M.~Olcese, A.~Torazza, A.~Trovato for their valuable support to T1;

- O.~Borysov, O.~Chykalov, Y.~Kostyshyn, M.~Protsenko, I.~Rosenko, O.~Starchenko and  I.~Tymchuk for their valuable support in assembling the GEM detectors of T2;

- M.G.~Bagliesi, R.~Cecchi, F.~Gherarducci, T.~Lomtadze, A.~Tazzioli for the design of electronics components and their help during the installation;

- M.~Besta, D.~Bochniak, E.~Denes, L.~Faber, P.~Janik, M.~Leinonen, T.~Palus, A.~Ster, K.~Ylilammi, M.~Wrobel, P.~Wyszkowski for their software development work;

- C.~Da Via, A.~Kok, S.~Parker, C.~Kenney, S.~Watts for the collaboration on 3D silicon detectors;

- the TE-CRG group, in particular F.~Haug and J.~Wu, for detector cooling developments;

- M.~A.~Ciocci, W.~Klempt, J.~Lippmaa, J.~Mill, G.~Notarnicola, G.~Rella, G.~Sanguinetti, F.~Sauli, S.Weisz for their contributions at earlier stages of the project.

We congratulate the CERN accelerator groups for the very successful operation in 2010.~We thank M.~Ferro-Luzzi and the LHC machine coordinators for scheduling the dedicated TOTEM fills.

We would like to acknowledge the crucial contributions from members of the following LHC groups:

- the LHC collimation group (R.~Assmann, R.~Bruce, S.~Redaelli, G.~Valentino, D.~Wollmann) for transferring
experience and techniques from collimator operation to the Roman Pot system;

- radiofrequency group (F.~Caspers, T.~Kroyer), F.~Roncarolo and A.~S\'{o}t\'{e}r for impedance studies;

- the beam instrumentation group (R.~Appleby, B.~Dehning, A.~Nordt, B.~Holzer, R.~Jones, M.~Sapinski) for their help with the BPMs and BLMs;
- the Machine Protection Panel (B.~Puccio, R.~Schmidt, J.~Wenninger, M.~Zerlauth) for the help on the interlock system;

- the beam optics group (H.~Burkhardt, M.~Giovannozzi, A.~Verdier) for the development of special optics.

We have to acknowledge also the valuable collaboration with PNPI, St.~Petersburg, for the T1 CSC assembly and tests and the help of A.~Fetisov during the test beam runs and the installation of T1 in IP5.

We are grateful to the INFN Scientific Committee I, in particular to the President F.~Ferroni and to the INFN Totem Referees, for the continuous and valuable support.

Furthermore, we thank CMS for the fruitful and effective collaboration and for providing their luminosity measurements.

This work was supported by the institutions listed on the front page and partially also by NSF (US), the Magnus Ehrnrooth foundation (Finland), the Waldemar von Frenckell foundation (Finland), the OTKA grant NK 73143 (Hungary) and by the NKTH-OTKA grant 74458 (Hungary).



\end{document}